\documentclass[a4paper,12pt]{article}
\makeatletter
\newcommand*{\length}[1]{%
    \@tempcnta\z@
    \@for\@tempa:=#1\do{\advance\@tempcnta\@ne}%
\the\@tempcnta
}
\makeatother

\usepackage{amssymb,amsmath,amsthm,mathrsfs,mathtools,bm,bbm,dsfont,relsize}
\usepackage{physics}
\usepackage{calc,float,caption,subcaption}
\usepackage{cite,graphicx,xcolor,datetime}
\usepackage[affil-it]{authblk}
\usepackage[symbol]{footmisc}
\usepackage{enumerate, array}
\usepackage{hyperref}

\usepackage{tikz,pgfplots,tkz-euclide,tikz-3dplot}
\usetikzlibrary{patterns,arrows,positioning,calc,fadings,decorations.pathreplacing,decorations,angles,quotes,intersections}
\pgfplotsset{compat=1.18}
\usepgfplotslibrary{colorbrewer}

\tikzstyle{block} = [rectangle, draw, fill=black!30,
    text width=6.3em, text centered, rounded corners, minimum height=3em]
\usepackage{geometry}
\newcommand\cone[1]{\operatorname{cone}(#1)}

\newcommand\Rf[1]{\widehat{#1}}
\newcommand\Ff[1]{\check{#1}}
\newcommand\G[1]{\Gamma\left(#1\right)}

\newcommand{\eq}[1]{~(\ref{#1})}

\newcommand\diag{\operatorname{diag}}

\newcommand\uvec[1]{\check{#1}}
\def\pa{\partial}

\def\tu{\tilde{u}}
\def\tphi{\tilde{\phi}}
\def\rt{\longrightarrow}

\def\dX{d^{d+1}X}
\def\dU{d^{d}U}
\def\D{\Delta}
\def\R{\mathbf{R}}
\def\Z{\mathbf{Z}}

\def\M{\mathcal{M}}
\def\Md{\M^{1,d}}
\def\dM{\partial\Md}
\def\eadsup{\M^{\uparrow}_-}
\def\eadsdn{\M^{\downarrow}_-}
\def\ds{\M_+}
\def\eads{\M_-}
\def\lcup{\M^{\uparrow}_0}
\def\lcdn{\M^{\downarrow}_0}
\def\lc{\M_0}
\def\Uds{\mathcal{U}_+}
\def\Uup{\mathcal{U}^\uparrow_-}
\def\Udn{\mathcal{U}^\downarrow_-}
\def\Ueads{\mathcal{U}_-}
\def\UzUp{\mathcal{U}^{\uparrow}_0}
\def\UzDn{\mathcal{U}^{\downarrow}_0}
\newcommand\sph[1]{\mathbf{S}^{#1}}
\def\I{\mathcal{I}}
\def\u{\vec{u}}
\def\ba{{\boldsymbol a}}
\def\bsigma{\Bar{\sigma}}
\def\sA{A_{\sigma}}
\def\bA{A_{\Bar{\sigma}}}
\def\m{\widehat{\boldsymbol\alpha}}
\newcommand\pFq[3]{
\ {}_{\length{#1}}F_{\length{#2}}%
\left(\left.\genfrac{.}{.}{0pt}{1}{#1}{#2}\right| {#3}\right)
}

\title{Minkowski Space holography and Radon transform}
\author{Samrat Bhowmick\thanks{email: bhowmicksamrat@gmail.com}~} 
\affil{\normalsize Department of Physics, 
\authorcr Savitribai Phule Pune University,
\authorcr Pune-411007}
\author{Koushik Ray\thanks{email: koushik@iacs.res.in}}
\affil{\normalsize Indian Association for the Cultivation of Science,
\authorcr Kolkata 700 032. India.}
\date{}

\begin{document}
 \begin{titlepage}
 \maketitle
\thispagestyle{empty}
 
\begin{abstract}
We relate a free scalar field in the Minkowski spacetime with a scalar 
field with a certain scaling dimension on a sphere of codimension two.
This is realised by first performing a Radon transform of the  ``bulk" field on 
the Minkowski space to a field on a hyperplane identified with
a de Sitter or  Euclidean anti-de Sitter slice. The Radon transform 
is identified in turn with a scalar field obtained from a sphere of
one further lower dimension through the so-called bulk reconstruction 
programme. 
We write down the Mellin modes of the bulk field as 
generalised hypergeometric functions using the Lee-Pomeransky method 
developed for evaluation of Feynman loop diagrams. 
\end{abstract}
\end{titlepage}
\section{Introduction}
\label{sec:intro}
The principle of  holography propounds describing
a quantum theory of gravity in a spacetime as a
quantum field theory sans gravity on some appropriately defined
boundary of the spacetime by relating the degrees of freedom of the
two theories \cite{tHooft:1993dmi, Susskind:1994vu}. 
It has been most concretely formulated as a duality with the 
type-IIB string theory as the theory of quantum gravity on a spacetime
containing the constant-curvature anti-de Sitter (adS) space within,
and the $N=4$ super Yang-Mills theory on its boundary
\cite{Maldacena:1997re, Witten:1998qj, 
Gubser:1998bc}. Several aspects of the duality have been generalised to 
the de Sitter (dS) space  as well
\cite{Strominger:2001pn, Witten:2001kn}. A similar formulation for the 
more mundane Minkowski spacetime is, however, only partway understood. 
The earliest attempts towards a ``flat-space holography" were based on 
the asymptotic behaviour of fields in the Minkowski space, taking into 
account the asymptotic symmetries 
\cite{deBoer:2003vf, Solodukhin:2004gs, Arcioni:2003xx, Barnich:2009se, 
Barnich:2010eb}. A  major hurdle in associating a dual conformal field theory 
to the Minkowski spacetime is the nonexistence of a temporal boundary. The 
presumptive dual theory is exiled to live on a sphere,
called the celestial sphere, at the  null infinity of the Minkowski spacetime 
\cite{Strominger:2017zoo,Banerjee:2018gce, Laddha:2020kvp, Iacobacci:2022yjo, 
Laddha:2022nmj, Sleight:2023ojm, Iacobacci:2024nhw,sugi}. 
Thus, the holographic duality is sought between theories on spaces 
whose dimensions differ by two. The framework, however,
successfully relates bulk scattering amplitudes in the 
Minkowski space to correlation functions of conformal fields living on the 
celestial sphere in accordance with the structure of asymptotic symmetry groups \cite{Barnich:2013jla,Strominger:2013lka,He:2014cra,Campiglia:2015qka,
Kapec:2015ena,He:2015zea}.
This is achieved by means of the Mellin transform of the external momenta, 
mapping momentum eigenstates in the Minkowski spacetime to operators with 
definite scaling dimensions and spins under the Lorentz group, which is also
the global conformal group on the celestial sphere 
\cite{Pasterski:2016qvg, Pasterski:2017kqt}. The celestial framework has 
emerged as a promising avenue for realising flat space holography and 
understanding the infrared structure of gauge and gravitational theories 
through the lens of asymptotic symmetries and soft theorems 
\cite{Laddha:2020kvp, Pasterski:2021rjz}. Approaches based on boundary 
conformal field theory and branes have
also been proposed \cite{Neuenfeld:2025wnl, taka}.

In here, we relate fields in the Minkowski spacetime, 
henceforth referred to as the bulk, to fields with a certain scaling 
dimension on a sphere of two less 
dimensions. As a first step, we only consider free, but massive,
scalar fields in the bulk. Such a field relates to Carrollian fields in the
far future \cite{kevin}.
The theories being non-gravitational,
the  association, while may be looked upon as a duality,  
does not qualify being a holography in the 
strict sense of the term. We, however, continue to use the term for lack of
a better one. On the other hand, the correspondence, being carried out using
the Radon transform, is invertible in a certain sense. 

Let us present a pr\'ecis of the association at the outset to be expounded 
in the rest of the article. The Minkowski bulk is treated as a foliation by 
hyperplanes, which are taken as either the Euclidean anti-de Sitter (EadS) 
or the dS space, depending on the region relative to the light cone, 
as depicted in Figure~\ref{fig1}. The broad idea is to usufruct the known 
descriptions of holography for EadS or dS spaces. A conformal field theory 
on the appropriately defined spherical boundary of these spaces is related 
to field theories on them through a procedure known as bulk reconstruction,
a notion inverse to the adS/CFT correspondence
\cite{Balasubramanian:1999ri, Bena:1999jv, hkll, Hamilton:2006az, 
Hamilton:2006fh, Kabat:2012hp, Kabat:2013wga, Kabat:2015swa, Roy:2015pga, 
Kabat:2017mun, Sanches:2017xhn, Goto:2017olq, Roy:2017hcp, Sarosi:2017rsq, 
Bhowmick:2017gj, Bhowmick:2018hie, Bhowmick:2019nso,balog}. 
This field theory is then related to one in
the Minkowski space through Radon transform. 
We consider a massive scalar field obeying 
the Laplace equation \eq{X:Lap} in the bulk. The Radon transform of 
the field defines a field on a family of hyperplanes \eq{u1} and 
satisfies the equation of a simple harmonic oscillator 
along the evolution of the family \eq{eq:fR}, the spring constant 
being real or
imaginary depending on whether the hyperplanes are in
the EadS or the dS space, \eq{unit:ads} or \eq{unit:ds}, respectively.
This allows for a separation of variables of the Radon
transform of the field into a part depending on the direction of evolution
and another depending on the coordinates
of the unital hyperplane of the family \eq{f:expand}. 
The hyperplanes being EadS or dS, we then identify the field 
on it as one obtained from the corresponding 
boundary, as in \eq{blk:rc}, through bulk reconstruction,
which in turn can be written as an
inverse Radon transform of a field with a certain scaling dimension on
a sphere \cite{Bhowmick:2017gj,Bhowmick:2019nso}. A further inversion 
of the Radon transform finally yields the bulk field \eq{phi:fin}.
This relates the massive free scalar field in the bulk to a field
with a certain scaling dimension on a sphere of two less dimensions as 
an integral transform, as described in the flow-chart below. 
\begin{figure}[h]
\begin{center}
\begin{tikzpicture}[node distance = 2.5in, auto]
\node[block] (B) at (0,0) {\small Field on hypersurface};
\node[block,left of= B] (A)  {\small Field on Minkowski space $\Md$};
\node[block,right of=B] (C) {\small Field on $\sph{d-1}$};
\draw[thick] ([yshift=.7em] A.east) edge[-latex] node[anchor=center,midway,above] 
{\tiny Radon transform} ([yshift=.7em] B.west);
\draw[thick] ([yshift=-.7em] B.west) edge[-latex] node[anchor=center,midway,below] 
{\tiny Inverse Radon transform} ([yshift=-.7em] A.east); 
\draw[thick,-latex] (C)  -- (B)  node [anchor=center,midway,above] {\tiny
Bulk reconstruction} node[anchor=center,midway,below] {\tiny
Inverse Radon transform};
\end{tikzpicture}
\end{center}
\end{figure}
It turns out that the Mellin modes of the massive
field in the bulk have a particularly 
nice analytic form as generalised GKZ hypergeometric
functions \eq{Mf:eads} and \eq{Mf:ds}. The integrals are explicitly performed
using the method of Lee-Pomeransky.

The organisation of the article is as follows. In the next section we present
the slicing of the Minkowski space into different regions and their
boundaries. We describe the 
patches covered by the EadS and the dS foliations and write out the 
coordinate charts explicitly in these patches. In section \ref{sec:Radon} 
we recall the definition and various properties of Radon transform
of scalar fields on the Minkowski space \cite{Japs}. 
We describe the relation between Radon and Fourier
transforms and use it to define the inverse Radon transform. We also 
show that the Radon transform of a free scalar field obeys an ``oscillator"
equation, with real and imaginary masses for the EadS and dS patches,
respectively. In section \ref{sec:hol} we lay out the holographic correspondence
by first separating variables using the oscillator equation and then relating
part of the field to one obtained through bulk reconstruction. 
In section \ref{sec:massless} we show that in the limit of vanishing mass 
the procedure matches the boundary value problem for the Laplace equation
in the Minkowski space. Finally, in section \ref{sec:rep} we indicate
the relationship of the integral geometry approach adapted here with 
representation theory of the proper Lorentz group on the Minkowski space 
and the conformal group of the sphere, two dimensions lesser, before 
summarising in section \ref{sec:summary}.
We write explicit formul{\ae} for the Mellin modes of the bulk field 
by performing integrations using the Lee-Pomeransky method developed 
for evaluation of Feynman loop diagrams, which is detailed in the Appendix. 
\begin{figure}[h]
\begin{center}
\begin{tikzpicture}
\begin{axis}[axis lines = center, ticks=none, 
view={15}{10},
domain=0:70,
    y domain=0:2*pi,
    xmin=-100,
    xmax=100,
    ymin=-100,
    ymax=100,
    zmin=-600,
    zmax=750,
zlabel={$\scriptstyle X^0$},
every axis z label/.style={at={(rel axis cs:0.4,0.5,1.05)},anchor=west},
samples=30
]
\addplot3[surf,colormap/PiYG
]({x*cos(deg(y))},{x*sin(deg(y))},{10*sqrt((0.7*x)^2+50)}); 
\addplot3[surf,domain=0:70,y domain=0:2*pi,colormap/PiYG
]({x*cos(deg(y))},{x*sin(deg(y))},{-10*sqrt((0.7*x)^2+50)}); 
\addplot3[surf,opacity=.2,domain=0:80,y domain=.83*pi:2.34*pi,colormap/YlOrRd
]({x*cos(deg(y))},{x*sin(deg(y))},{4*x}); 
\addplot3[surf,opacity=.2,domain=0:80,y domain=2*pi:0,colormap/YlOrRd
]({x*cos(deg(y))},{x*sin(deg(y))},{-4*x}); 
\addplot3[surf,opacity=.2,domain=-35:35,y domain=0:2*pi
]({3*sqrt(50+x*x)*cos(deg(y))},{3*sqrt(50+x*x)*sin(deg(y))},{9*x})
; 
\addplot3[point meta=explicit symbolic,nodes near coords] 
coordinates {(97,88,200)[$\scriptscriptstyle\ds$]};
\addplot3[point meta=explicit symbolic,nodes near coords] 
coordinates {(25,90,420)[$\scriptscriptstyle\eadsup$]};
\addplot3[point meta=explicit symbolic,nodes near coords] 
coordinates {(25,10,-600)[$\scriptscriptstyle\eadsdn$]};
\addplot3[point meta=explicit symbolic,nodes near coords] 
coordinates {(75,10,305)[$\scriptscriptstyle\lcup$]};
\addplot3[point meta=explicit symbolic,nodes near coords] 
coordinates {(75,0,-370)[$\scriptscriptstyle\lcdn$]};
\end{axis}
\end{tikzpicture}
\end{center}
\caption{Different regions of $\Md$
\label{fig1}}
\end{figure}
\section{Slicing the  Minkowski space}
In this section we describe the geometrical setup. 
Let $\Md$ denote the $(d+1)$-dimensional  
Minkowski spacetime with a pseudo-Riemannian metric
$\eta=\diag (-1,1,\cdots,1)$. 
Let $X=\{X^{\mu}\}=\{(X^0,X^i, X^d)\}$, with $i=1,2,\cdots,d-1$, 
denote a vector in $\Md$ with norm $|X|$, defined by 
\begin{equation}
\label{Mnorm}
|X|^2=\eta_{\mu\nu}X^{\mu}X^{\nu}=-(X^0)^2+(X^1)^2+\cdots+(X^{d-1})^2+ (X^d)^2.
\end{equation} 
The inner product of two vectors $X$ and $Y$ in $\Md$ is  
\begin{equation}
\label{etamn}
\begin{split}
     X\cdot Y &= \eta_{\mu\nu}X^{\mu}Y^{\nu}\\
     &= -X^0Y^0+X^1Y^1+\cdots + X^{d-1}Y^{d-1}+X^dY^d.
\end{split}
   \end{equation}
The spacetime is partitioned into different regions or patches, 
\begin{align}
	\ds = \{X\in\Md ;\ |X|^2 > 0\}, &\\
\begin{gathered}
	\eadsup= \{X\in\Md; |X|^2 < 0, \ X^0>0, \},\\
	\eadsdn = \{X\in\Md; |X|^2 < 0,\ X^0<0, \},\\
\end{gathered}
		& \qquad\eads=\eadsup\cup\eadsdn,\\
\begin{gathered}
	\lcup = \{X\in\Md; |X|^2= 0,\ X^0>0 \}, \\
	\lcdn = \{X\in\Md; |X|^2 = 0,\ X^0<0 \}, 
\end{gathered}
& \qquad\lc = \lcup\cup\lcdn
\end{align}
diagrammatically presented in Figure~\ref{fig1}.
The partitioning depends on the sign of the 
squared norm defined by \eq{Mnorm}, indicated in the 
subscript of $\mathcal{M}$, 
and the sign of the temporal coordinate of a point,  
indicated in the superscript. 
Along with the origin, these cover the whole Minkowski space,
namely,
\begin{equation}
\label{MinCover}
    \Md = \ds\cup\eads\cup\lc\cup\{0\}.
\end{equation}
For our purposes, these different regions are to be described as foliations 
by hyperplanes. Let us define sets of unit vectors 
whose dilations cover different parts of \eq{MinCover}. 
\begin{align}
    \Uds =\{X\in\Md;   |X|^2=1\}, &\\
\begin{gathered}
\Uup = \{X\in\Md; |X|^2=-1, X^0>0\},  \\
\Udn = \{X\in\Md; |X|^2=-1, X^0<0\}, 
\end{gathered}
&\qquad\Ueads =\Uup\cup\Udn,\\
   \UzUp = \{X\in\Md; |X|^2=0, X^0=1\},  &\\
    \UzDn = \{X\in\Md; |X|^2=0, X^0=-1\}.&
    \
\end{align}
We refer to $\Uds$ as the de Sitter (dS) space and $\Ueads$ as the 
Euclidean anti-de Sitter (EadS) space.
The set of spacelike vectors in 
$\ds$, called the de Sitter (dS) patch, is the dilation of $\Uds$, written as 
\begin{equation}
\ds = \varrho\ \Uds := \{X=\varrho U; \varrho\neq 0, U\in\Uds\}. 
\end{equation}
Similarly, the region $\eads$ is called the EadS patch. 
The two pieces $\eadsup = \varrho\ \Uup$ and $\eadsdn=\varrho\ \Udn$, 
with $\varrho>0$ are called the upper and lower lobe, respectively.
The regions $\lcup$ and $\lcdn$ are, respectively, 
the forward and backward light 
cones, with their union $\lc$ the light cone.
Equation \eq{MinCover} can thus be rewritten as
\begin{equation}
\Md = \underset{\varrho\neq 0}{(\varrho\,\Uds)} \cup\underset{\varrho>0}{%
(\varrho\,\Uup)}\cup\underset{\varrho>0}{(\varrho\,\Udn)}\cup\lcup\cup\lcdn\cup\{0\}.
\end{equation}
A holographic scheme requires  a boundary, but the Minkowski space is 
open. The space 
\begin{equation}
    \dM = \Uds\cup\Uup\cup\Udn\cup\UzUp\cup\UzDn
\end{equation}
is taken to be the boundary by proxy \cite{Japs,helgabook}. 
The Radon transform takes a bulk field to hyperplanes.
For a point $X$ in $\Md$ and a point $U$ in $\dM$, a hyperplane
$\varpi(\lambda,U)$ is defined by 
\begin{equation}
\label{u1}
\lambda+U\cdot X =0,
\end{equation}
where $\lambda$ is a real number. 
The set of all such hyperplanes is denoted $\Pi$. Varying $\lambda$ 
yields a family of hyperplanes in $\Pi$.
The hyperplanes thus defined satisfy   
 $\varpi(\lambda,U)=\varpi(-\lambda,-U)$ and 
$\varpi(0,\varrho U)=\varpi(0,U)$, for $U\in\dM$ and $\varrho\in\R$.
It follows that the parameter space of the space of hyperplanes
$\Pi$ is $\Uup\cup(\Uds/\Z_2)\cup\UzUp$, where 
$\Z_2=\{-1,+1\}$. Let us introduce coordinate charts on the two patches. 
\subsection{EadS patch}
The EadS patch $\eads$ is covered with coordinates
\begin{equation}
\begin{gathered}
\label{patch:eads}
X^0 = \frac{\xi \tau}{2}  \left(1+\frac{1+x^2}{\tau^2}\right), \quad X^i = \frac{x^i \xi}{\tau},  \quad  X^d = \frac{\xi \tau}{2}  \left(1-\frac{1-x^2}{\tau^2}\right),\\
-\infty < \xi < \infty,\quad 0\leqslant\tau< \infty,\quad -\infty < x^i < \infty ,
\end{gathered}
\end{equation}
where $i=1,2,\cdots,d-1$ 
and we write $x^2=\sum_{i=1}^{d-1} (x^i)^2$. 
The norm of $X$ in this patch is $|X|^2=-\xi^2$.
The metric $\eta$ is given by the line element
\begin{equation}
 ds^2 = -d\xi^2 + \frac{\xi^2}{\tau^2} (d\tau^2 + dx^2)
\end{equation}
in the EadS patch
with $\xi$ as the timelike coordinate. Here 
$\xi>0$ corresponds to the upper lobe $\eadsup$ and 
$\xi<0$ to the lower lobe $\eadsdn$, and $\xi=0$ is the origin.

The volume element in this chart is 
\begin{equation}
	\label{vol2}
     \dX = \qty(\frac{\xi}{\tau})^{d} d\xi d\tau d^{d-1}x.
\end{equation}
where $d^{d-1}x=dx^1dx^2\ldots dx^{d-1}$. 
%
A point in the boundary region $\Ueads$ is obtained by setting
$\xi=1$ in \eq{patch:eads}. Let us write it as
\begin{equation}
\begin{gathered}
\label{unit:ads}
 U^0 = \frac{t}{2}  \left(1+\frac{1+u^2}{t^2}\right), 
 \quad U^i = \frac{u^i}{t},  
 \quad U^d = \frac{t}{2}  \left(1-\frac{1-u^2}{t^2}\right),\\
  0\leqslant t< \infty,\quad -\infty < u^i < \infty ,
\end{gathered}
\end{equation}
leading to the hyperplane $\lambda+U\cdot X=0$, with 
\begin{equation}
\label{UX:eads}
    U\cdot X = -\frac{\xi}{2\tau t}((x-u)^2+\tau^2+t^2) 
\end{equation}
in this patch.

\subsection{dS patch}
The de Sitter patch  $\ds$ is covered with coordinates,
\begin{equation}
\begin{gathered}
\label{patch:ds}
 X^0 = \frac{\xi \tau}{2}  \left(1-\frac{1+x^2}{\tau^2}\right), \quad X^i = \frac{x^i \xi}{\tau},  \quad X^d = \frac{\xi \tau}{2}  \left(1+\frac{1-x^2}{\tau^2}\right),\\
\xi>0,\quad  -\infty<\tau<\infty,\quad -\infty < x^i < \infty.
\end{gathered}
\end{equation}
The norm of $X$ in this patch is $|X|^2=\xi^2$. 
The line element takes the form 
\begin{equation}
    ds^2  = d\xi^2 +\frac{\xi^2}{\tau^2} (-d\tau^2+dx^2),
\end{equation}
in this patch, with the timelike coordinate $\tau$.
The volume element once again assumes the form 
\begin{equation}
	\label{vol1}
     \dX = \qty(\frac{\xi}{\tau})^{d} d\xi d\tau d^{d-1}x.
\end{equation}

In order to  describe the hyperplanes in this patch
let us define a point $U\in\Uds$ of the boundary as 
\begin{equation}
\begin{gathered}
\label{unit:ds}
 U^0 = \frac{t}{2}  \left(1-\frac{1+u^2}{t^2}\right), 
 \quad U^i = \frac{u^i}{t},  
 \quad U^d = \frac{t}{2}  \left(1+\frac{1-u^2}{t^2}\right),\\
  -\infty < t<\infty,\quad -\infty < u^i < \infty.
\end{gathered}
\end{equation}
A hyperplane is then defined as $\lambda+U\cdot X=0$ with
\begin{equation}
\label{UX:ds}
    U\cdot X =-\frac{\xi}{2\tau t}((x-u)^2-\tau^2-t^2).
\end{equation}

In both the EadS and the dS space the volume element takes the form 
$\Uup$, namely,
\begin{equation}
\label{Uvol}
    \dU = \frac{dtd^{d-1}u}{t^d}.
\end{equation}
Sometimes we write $U=(t,u)$ to indicate the variables explicitly.
\section{Radon transform}
\label{sec:Radon}
The holographic correspondence is formulated in terms of 
Radon transform of fields on the Minkowski space. 
In this section we discuss 
certain features of Radon transform \cite{Japs,helgabook,gelfand} of 
scalar fields germane to our present purposes.
The Radon transform is an integral transform. 
In order to define it we need a normalised measure on $\Md$. The
normalisation is fixed by declaring that the field is supported solely on 
$\ds\cup\eads$. As mentioned above, given a point $X$ in $\Md$ in the EadS or
dS patch given by \eq{patch:eads} or \eq{patch:ds}, respectively, the 
corresponding boundary vector $\uvec{X}$ in $\dM$ is obtained by setting 
$\xi=\pm 1$. We write the point as the pair $X=(\xi,\uvec{X})$ in any
patch. The integral of a scalar 
field, $\phi(X)$ on $\Md$ is then defined as
\begin{equation}
\label{measure:def}
    \int\limits_{\Md}\ \dX \phi(X) = \int\xi^d d\xi 
	\int\limits_{\Uds\cup\ \Uup\cup\ \Udn}\frac{d\tau}{\tau^d}\, 
	d^{d-1}x\  \phi(\xi,\uvec{X}),
\end{equation}
where $\dX$ is defined on the two patches in \eq{vol2} and \eq{vol1}. 
We use the notations $\phi(X)=\phi(\xi\uvec{X})=\phi(\xi,\uvec{X})$.
Integral transforms are defined with respect
to the measure thus normalised. 
Let us note further that for a local field theory, we deal with 
fields supported on only one of the regions $\ds$, $\eadsup$ or $\eadsdn$, 
and the range of integration over $\xi$ and $\tau$ are taken according 
to \eq{patch:eads}  and \eq{patch:ds}.

A function $\varphi(\varpi)$ on the set of hyperplanes $\Pi$ 
is taken to be a function $\varphi(\lambda,U)$, defined on $\R\times\dM$, 
with $U\in\dM$ and $\lambda\in\R$. 
The Radon transform of a scalar field $\phi$ is an integral transform 
which restricts the 
field to a hyperplane $\varpi(\lambda,U)$ defined by \eq{u1}, 
\begin{equation}
\label{Radon:def}
	\Rf{\phi}(\varpi)=\Rf{\phi}(\lambda,U)= \int\limits_{\varpi} d\varpi\ 
	\phi(X) = \int\limits_{\Md} \dX\ \phi(X)\delta(\lambda+U\cdot X),
\end{equation}
defining a function on the space of hyperplanes $\Pi$ with $d\varpi$ 
denoting the restriction of the measure $\dX$ on $\varpi$. 
This allows us to separate variables on the hyperplanes. 
Defining $\alpha = \lambda+U\cdot X$, we have, 
\begin{equation}
\frac{\partial}{\partial\lambda} \delta(\lambda+U\cdot X)
= \frac{d}{d\alpha} \delta(\lambda+U\cdot X).
\end{equation}
Then, 
\begin{equation}
\label{Xl}
\begin{split}
\frac{\partial}{\partial X^{\mu}} \delta(\lambda+U\cdot X)
&= U_{\mu}\frac{d}{d\alpha} \delta(\lambda+U\cdot X)\\
&=U_{\mu}\frac{d}{d\lambda} \delta(\lambda+U\cdot X),
\end{split}
\end{equation}
where $\mu=0,1,\ldots,d$.
Let
\begin{equation}
\Box = \eta^{\mu\nu}\frac{\pa^2}{\pa X^{\mu}\pa X^{\nu}}
\end{equation}
be the pseudo-Laplacian with the metric used in \eq{etamn}. Then, the Radon
transform of $\Box\phi$ becomes
\begin{equation}
\begin{split}
\Rf{\Box\phi}(\lambda,U) 
&= \int\limits_{\Md} \dX \Box\phi(X) \delta(\lambda+ U\cdot X)\\
&= -\eta^{\mu\nu} \int\limits_{\Md} \dX \frac{\pa\phi(X)}{\pa X^{\mu}}
\frac{\pa}{\pa X^{\nu}}\delta(\lambda+U\cdot X)\\
&= -U_{\mu} \int\limits_{\Md} \dX \frac{\pa\phi(X)}{\pa X^{\mu}}
\frac{d}{d\lambda}\delta(\lambda+U\cdot X),
\end{split}
\end{equation}
where we integrated by parts in the first step and used \eq{Xl} in the second. 
Integrating by parts and using \eq{Xl} once again along with \eq{Radon:def},
we then obtain
\begin{equation}
\label{LapLam}
\Rf{\Box\phi}(\lambda,U)  = |U|^2\frac{d^2\Rf{\phi}(\lambda,U)}{d\lambda^2},
\end{equation}
with $|U|^2=\pm 1$ in the patches \eq{patch:eads} and \eq{patch:ds}.

An inversion formula for the Radon transform 
of a field is obtained through its Fourier transform. 
For a vector $k\in\Md$ expressed as $k=pU$, with $p\in\R\setminus\{0\}$ 
and  $U\in\dM$, the Fourier transform of  a scalar field in $\Md$ is 
\begin{equation}
\label{Fou}
\begin{split}
\Ff{\phi}(k) 
&= \Ff{\phi}(pU) \\
&= \int\limits_{\Md} \dX e^{-ipU\cdot X} \phi(X)\\
&= \int\limits_{-\infty}^{\infty} d\lambda 
\int\limits_{U\cdot X =-\lambda} d\varpi\
e^{ip\lambda} \phi(X)\\
&=  \int\limits_{-\infty}^{\infty} d\lambda e^{ip\lambda} 
\Rf{\phi}(\lambda,U).
\end{split}
\end{equation}
Thus, the $(d+1)$-dimensional Fourier transform is the one-dimensional 
Fourier transform of the Radon transform. 
The inversion formula is derived by first writing the field as the inverse 
Fourier transform of $\Ff{\phi}$ 
\begin{equation}
\label{IFou}
\phi(X) = \left(\frac{1}{2\pi}\right)^{d+1} \int\limits_{\Md} d^{d+1}k
\Ff{\phi}(k) e^{ik\cdot X},
\end{equation}
followed by using \eq{Fou}, to derive 
\begin{equation}
\label{IRadon}
\begin{split}
\phi(X) 
&= \left(\frac{1}{2\pi}\right)^{d+1} 
\int\limits_{\dM} \dU \int\limits_{-\infty}^{\infty} dp\ p^d
\int\limits_{-\infty}^{\infty} d\lambda\
e^{ip(U\cdot X+\lambda)} \Rf{\phi}(\lambda,U) \\
&= \left(\frac{1}{2\pi}\right)^{d+1} 
\int\limits_{\dM} \dU \int\limits_{-\infty}^{\infty} d\lambda
\int\limits_{-\infty}^{\infty} dp\
e^{ip(U\cdot X+\lambda)} \left(\frac{id}{d\lambda}\right)^d
\Rf{\phi}(\lambda,U) \\
&= \left(\frac{i}{2\pi}\right)^{d} 
\int\limits_{\dM} \dU \int\limits_{-\infty}^{\infty} d\lambda
\delta(U\cdot X+\lambda)
\left(\frac{d^d\Rf{\phi}(\lambda,U)}{d\lambda^d}\right) \\
&= \left(\frac{i}{2\pi}\right)^{d} 
\int\limits_{\dM} \dU 
\left(\frac{d^d\Rf{\phi}(\lambda,U)}{%
d\lambda^d}\right)_{\lambda=-U\cdot X}, 
\end{split}
\end{equation}
where we have used integration by parts in the second step. This defines
the inverse of the Radon transform.
\section{Holographic correspondence}
\label{sec:hol}
We now proceed to discuss the holographic correspondence. 
Let us consider a free scalar field in the Minkowski space.
A free scalar field $\phi(X)$ satisfies the Laplace equation
\begin{equation}
\label{X:Lap}
\Box \phi(X)=m^2 \phi(X)
\end{equation}
in the bulk Minkowski space $\Md$. Taking Radon transform on both sides, 
and using \eq{LapLam}, the Radon transform of the field satisfies the equation
\begin{equation}
\label{eq:fR}
\frac{\partial^2}{\partial\lambda^2}\Rf{\phi}(\lambda,U)
=m^2|U|^2\Rf{\phi}(\lambda,U),
\end{equation}
which is solved to obtain an expression for the 
field $\Rf{\phi}$ on $\varpi(\lambda,U)$ as
\begin{equation}
\label{f:expand}
\Rf{\phi}(\lambda,U) = e^{m\lambda\,\sqrt{|U|^2}} \phi_1(U)
+e^{-m\lambda\,\sqrt{|U|^2}}\phi_2(U),
\end{equation}
where $\phi_1$ and $\phi_2$ are arbitrary functions on $\dM$.
Constants of integration are absorbed in the $\phi$'s.
Let us emphasize the importance of using the Radon transform at this point.
It leads to a separation of variables for the 
transformed field on hyperplanes in terms of $\lambda$ and $U$.
We can take the arbitrary fields $\phi_1$ and $\phi_2$ to be supported on the
EadS space $\Ueads$ or the dS space $\Uds$. 
Plugging \eq{f:expand} into the inversion formula \eq{IRadon} we then obtain 
\begin{equation}
\label{phi:mnk}
\phi(X) = \left(\frac{i}{2\pi}\right)^{d} \int\limits_{\dM} \dU \left(
(m\sqrt{|U|^2})^d e^{-m\sqrt{|U|^2} U\cdot X} \phi_1(U)
+(-m\sqrt{|U|^2})^d e^{m\sqrt{|U|^2} U\cdot X} \phi_2(U)
\right).
\end{equation}
Conversely,  given a field on the EadS or dS space we obtain a free field 
in the Minkowski space. This is seen by acting on both sides with the 
pseudo-Laplacian operator $\Box$ to derive the free field equation \eq{X:Lap}.
This is the first step in the holographic correspondence. The second step
is to relate the scalar field to a conformal field on a sphere
in two lower dimensions than the bulk. This is achieved
by assuming that the fields $\phi_1$ and $\phi_2$ are 
obtained from the boundary of the EadS or dS space through bulk reconstruction.
Denoting the coordinates on the boundary sphere $\sph{d-1}$ 
of the EadS or dS space as $\tu$, 
a field  $\phi(U)$ in the EadS or dS space is expressed in terms of an 
integral of a field $\tphi(\tu)$ with a kernel $K(U|\tu)$, so that the field 
$\Rf{\phi}$ is written as 
\begin{equation}
\label{blk:rc}
\Rf{\phi}(\lambda,U)= e^{m \lambda \sqrt{|U|^2}} 
\int\limits_{\sph{d-1}} K(U| \tilde{u}) \tilde{\phi}_1(\tilde{u}) \, 
d^{d-1} \tilde{u}
+ e^{-m \lambda \sqrt{|U|^2}} 
\int\limits_{\sph{d-1}} K(U| \tilde{u}) \tilde{\phi}_2(\tilde{u}) \, 
d^{d-1} \tilde{u}, 
\end{equation}
so that the bulk field \eq{phi:mnk} assumes the form 
\begin{multline}
\label{phi:fin}
\phi(X) = \left(\frac{i}{2\pi}\right)^{d} \int\limits_{\dM} \dU \left(
(m\sqrt{|U|^2})^d e^{-m\sqrt{|U|^2} U\cdot X} 
\int\limits_{\sph{d-1}} K(U| \tilde{u}) \tilde{\phi}_1(\tu) 
d^{d-1}\tu 
\right.\\\left.+(-m\sqrt{|U|^2})^d e^{m\sqrt{|U|^2} U\cdot X} 
\int\limits_{\sph{d-1}} K(U| \tilde{u}) \tilde{\phi}_2(\tu) d^{d-1}\tu
\right).
\end{multline}
The kernel $K$ is taken to be the HKLL kernel \cite{hkll} arising in the
bulk reconstruction programme. 
Let us now recall the bulk reconstruction obtaining the kernel 
in terms of the inverse of 
the horospherical Gel'fand-Graev-Radon (GGR) transform
from the boundary \cite{Bhowmick:2017gj,Bhowmick:2019nso}.
It appears convenient to discuss the two patches separately.
\subsection{Holography in the EadS patch}
\subsubsection{Bulk reconstruction in EadS}
\label{bulk:eads}
The Radon transform on the EadS space is given in terms of a null vector. For 
a vector $\rho$ on the positive light cone $\lcup$ a horosphere is 
given by $U\in\Ueads$ satisfying the equation
\begin{equation}
\label{hor:eads}
\rho\cdot U+1=0.
\end{equation}
The null cone $\lcup$ is a metric cone $\R_+\times_{\rho^0}\sph{d-1}$.
A null vector commensurate with the coordinates \eq{unit:ads} is given by 
\begin{equation}
\begin{gathered}
\label{xipara}
\rho^i = \frac{2 \tu^i}{1+\tu^2} \rho^0, 
\quad\rho^d = - \frac{1-\tu^2}{1+\tu^2} \rho^0\quad 
\tu^2= \sum_{i=1}^{d-1} (\tu^i)^2,\\
0\leqslant\rho^0 < \infty,\quad -\infty < \tu^i < \infty,
\quad i=1,\cdots,d-1.
\end{gathered}
\end{equation}
The volume element on $\lcup$ in this chart is
\begin{equation}
\label{dxi}
\begin{split}
d\rho &= \frac{d\rho^0\cdots d\rho^{d-1}}{\rho^d}\\
&= \frac{2^{d-1} (\rho^0)^{d-2}}{\left(1+\tu^2\right)^{d-1}} 
d\rho^0 \, d^{d-1} \tu.
\end{split}
\end{equation}
The GGR transform of a field $\varphi(U)$ on $\Ueads$ is defined as
\begin{equation}
\label{radon:eads}
\Rf{\varphi}(\rho) =\int_{\Ueads} \varphi(U) 
\delta\left(\rho\cdot U+1\right) \dU,
\end{equation}
where $\dU$ is defined in \eq{Uvol}. The inverse of the GGR transform, in 
turn, is given by
\begin{equation}
\label{irad}
\varphi(U) = c_d\int_{\lcup} 
\frac{\Rf{\varphi}(\rho)}{\left|\rho\cdot U+1\right|^d} d\rho,
\end{equation}
where $c_d$ is a constant, depending on the dimension $d$.
We define the field on the conformal boundary  as
$\tphi(\tu)=\Rf{\varphi}(\rho)$. Assuming further that on the
horosphere \eq{hor:eads}, the boundary field scales as
$\tphi(p\tu)=p^{-\Delta}\tphi(\tu)$, we have
\begin{equation}
\label{cft:assumads}
\begin{split}
\Rf{\varphi}(\rho) &= \Rf{\varphi}(\rho^0,\cdots,\rho^{d-1})\\
&=\tphi\left(
\frac{2 \tu^1}{1+\tu^2} \rho^0, 
\frac{2 \tu^2}{1+\tu^2} \rho^0, \cdots,
\frac{2 \tu^{d-1}}{1+\tu^2} \rho^0
\right) \\ 
&=\left(\frac{2\rho^0}{1+\tu^2}\right)^{-\Delta} 
\tphi(\tu).
\end{split}
\end{equation}
Inserting this in \eq{irad} and performing the integration over $\rho^0$
we obtain
\begin{equation}
\label{eads2cone}
\varphi(U)=
\phi_0(d,\Delta)\int_{\sph{d-1}}{K}(t,u|\tu)
\tphi(\tu) \, d^{d-1}\tu \;,
\end{equation}
where the kernel is
\begin{equation}
\label{smrfunceads}
{K}(t,u|\tu) = 
{\left(\frac{t}{t^2+(u-\tu)^2}\right)^{d-1-\Delta}},
\end{equation}
and $\phi_0(d,\Delta)$ is a constant,
\begin{equation}
\label{phi0:eads}
\phi_0(d,\Delta)=\frac{e^{id\pi/2}}{2^{\Delta-1}\pi^{(d-3)/2}}
\frac{\G{\Delta+1}}{\G{\Delta-d+2}\G{\frac{d+1}{2}}}.
\end{equation}
Identifying $\varphi(U)$ with $\phi_1(U)$ or $\phi_2(U)$ in \eq{f:expand}
we obtain  \eq{blk:rc} with the kernel \eq{smrfunceads}.
\subsubsection{Bulk field in EadS patch}
The free scalar field in the EadS patch \eq{patch:eads} is obtained 
as $\phi(X)=\phi(\xi,\tau,x)$ by setting $|U|^2=-1$, plugging in \eq{Uvol},
\eq{UX:eads} and \eq{smrfunceads} in \eq{phi:fin}, and performing the 
integration over $U=(t,u)$. 
The Mellin modes of the bulk field with respect to the family parameter 
$\xi$ of the foliation \eq{patch:eads}, expressed in terms of the fields on the
sphere can be expressed as generalised hypergeometric functions. 
In the EadS patch the domain of $\xi$ is the whole real line. However, it is 
reasonable to assume that the field is supported on either of the lobes of
\eq{measure:def}. Considering fields supported on the upper lobe 
$\eadsup$, the Mellin modes are
\begin{equation}
\label{Mphi:eads}
\phi_s(\tau,u)=\int\limits_0^\infty \xi^{s-1}d\xi \phi(\xi,\tau,x).
\end{equation}
Plugging $X$ from \eq{patch:eads} in \eq{phi:fin} and 
performing the integration over $\xi$ we obtain, 
\begin{multline}
\label{Mf:eads}
\phi_s(\tau,x) = 
\left(\frac{i}{2\pi}\right)^d 
(im)^{d-s} (2\tau)^s\G{s} 
\int_{\sph{d-1}} d^{d-1}\tu\tphi_1(\tu)
\I(+1,s,d-1-\Delta)\\
+\left(\frac{i}{2\pi}\right)^d 
(-im)^{d-s} (2\tau)^s\G{s} 
\int_{\sph{d-1}} d^{d-1}\tu\tphi_2(\tu)
\I(+1,s,d-1-\Delta),
\end{multline}
where the integral $\I$ is defined in \eq{master}. Plugging in the values
of $\epsilon$, $\alpha_1$ and $\alpha_2$,  the integral $\I$ is a linear 
combination of three terms as discussed in the Appendix, namely,
\begin{equation}
\label{lin:com}
\I = C_1(\I_{123}^{(1)} + \I_{123}^{(2)}) + C_2\I_{234},
\end{equation}
where $C_1$ and $C_2$ are arbitrary constants which are to be fixed by 
imposing desired boundary or asymptotic behaviour of the bulk field.
We write the  explicit expressions of the three terms below.  
\begin{multline}
\label{eads:I123:1}
\I_{123}^{(1)}(+1,s,d-1-\Delta)=
i^{d+1}
\sum_{n_1,n_2=0}^{\infty}
2^{\frac{\D-2-s-2n_1}{2}}
\\
\frac{%
\sec\frac{\pi}{2}(1-2d+s+\D-2n_1-2n_2)\sin\pi(d-1-\D+2n_1+n_2)
\G{\frac{d-1}{2}}\G{\frac{s-\D+2n_1}{2}}
}{%
\scriptstyle
\G{2-d+\frac{s+\D}{2}-n_1-n_2}
\G{\frac{d+1-2s}{2}+n_1+n_2}
\G{1+s-\D-n_2}
\G{d-1-\D}\G{\D-s}\G{s}
\G{1+n_1}\G{1+n_2}
}
\\
\tau^{d-1-2s+2n_1+2n_2}
((x-\tu)^2+\tau^2)^{1-d+\frac{s+\D}{2}-n_1-n_2}
\pFq{\frac{s-\D}{2}+n_1,\frac{1-d}{2}+s-n_1-n_2}{
{\frac{4-2d+s+\D-2n_1-2n_2}{2}}}{\frac{\tau^2+(x-\tu)^2}{2\tau^2}},
\end{multline}
\begin{multline}
\label{eads:I123:2}
\I_{123}^{(2)}(+1,s,d-1-\Delta)=
-i^{d+1}
\sum_{n_1,n_2=0}^{\infty}
2^{\D-d-2n_1-n_2}
\\
\frac{%
\cos\frac{\pi}{2} (d-2s+2n_1+2n_2)\sec\frac{\pi}{2}(1-2d+s+\D-2n_1-2n_2)
\G{\frac{d-1}{2}}
\G{\frac{d-1+s-\D}{2}}
}{%
\scriptstyle
\G{2-d+\D-2n_1-n_2}
\G{d-\frac{s+\D}{2}+n_1+n_2}
\G{1+s-\D-n_2}
\G{d-1-\D}\G{\D-s}\G{s}
\G{1+n_1}\G{1+n_2}
}
\\
\tau^{1-d-s+\D}
\pFq{\frac{d-1-\D+s}{2},d-1-\D+2n_1+n_2}{d-\frac{s+\D}{2}+n_1+n_2}{
\frac{\tau^2+(x-\tu)^2}{2\tau^2}},
\end{multline}
\begin{multline}
\label{eads:I234}
\I_{234}(+1,s,d-1-\Delta)=
-i^{d+1}
\sum_{n_1,n_2=0}^{\infty}
2^{\frac{d-3-3s+\D}{2}-n_1+n_2}
\\
\frac{%
\csc\frac{\pi}{2}(d-1+s-\D+2n_1)\sin\pi(1-2s+\D-2n_1+n_2)
\G{\frac{d-1}{2}}
\G{\frac{1-d+3s-\D+2n_1-2n_2}{2}}
}{%
\scriptstyle
\G{\frac{3-d-s+\D-2n_1}{2}}
\G{\frac{3-d}{2}+s+n_1-n_2}
\G{1+s-\D-n_2}
\G{d-1-\D}\G{\D-s}\G{s}
\G{1+n_1}\G{1+n_2}
}%
\\
\tau^{2n_1}
(\tau^2+(x-\tu)^2)^{\frac{1-d-s+\D-2n_1}{2}}
\pFq{-n_1,\frac{1-d+3s-\D+2n_1-2n_2}{2}}{\frac{3-d-s+\D-2n_1}{2}}{
\frac{\tau^2+(x-\tu)^2}{2\tau^2}}.
\end{multline}
\subsection{Holography in the dS patch}
The construction in the dS space runs parallel to the EadS case. 
\subsubsection{Bulk reconstruction in dS}
\label{bulk:ds}
A horosphere in the dS space is defined as the hypersurface
\begin{equation}
\label{hor:ds}
|\rho\cdot U|=1,
\end{equation}
where $U\in\Uds$, and $\rho$ is a null vector in $\Md$, 
that is, $\rho\in\lc$. The light
cone is a metric cone $\R_+\times_{\rho^0}\sph{d-1}$ over the 
$(d-1)$-dimensional sphere. The affine chart on it commensurate with 
\eq{unit:ds} is 
\begin{equation}
\begin{gathered}
\label{xiparads}
\rho^i = -\frac{2 \tilde u^i}{1+\tilde{u}^2} \rho^0, 
\quad\rho^d = - \left(\frac{1-\tilde{u}^2}{1+\tilde{u}^2}\right)\rho^0,\quad
\tilde{u}^2= \sum_{i=1}^{d-1} (\tilde{u}^i)^2,\\
-\infty <\rho^0 < \infty,
\quad-\infty < \tilde{u}^i < \infty,\quad i=1,\cdots,d-1.
\end{gathered}
\end{equation}
In this chart we have 
\begin{equation}
\label{xidotX:ds}
\rho\cdot U=
\frac{\rho^0\big(t^2+(u-\tilde u)^2\big)}{t \left(1+\tilde u^2\right)}.
\end{equation}
The conformal boundary is given by $\rho\cdot U=0$, situated at $t\rt0_{\pm}$
and $u\rt\tilde{u}$ in the affine chart. 
Choosing either sign corresponds to the quotient $\Uds/\Z_2$
alluded to above.
The GGR transform of a scalar field $\varphi(U)$ in $\Uds$ is defined as
\begin{equation}
 \label{Radonds}
\Rf{\varphi}(\rho)=\int_{\Uds} \varphi(U) 
\delta\left(|\rho\cdot U|-1\right) \dU,
\end{equation}
where $\dU$ is defined in \eq{Uvol}. The inverse of the GGR transform is
\begin{equation}
 \label{iradds}
\varphi(U) = c_d\int_{\lc} 
\frac{\Rf{\varphi}(\rho)}{\left(|\rho\cdot U|- 1\right)_+^d} d\rho,
\end{equation}
where  $x_+^a=\theta(x)x^a$, with $\theta(x)$ denoting the Heaviside step
function,  and the volume element 
\begin{equation}
\label{dxids}
\begin{split}
d\rho &= \frac{d\rho^0\cdots d\rho^{d-1}}{\rho^d}\\
&=\frac{2^{d-1} (-\rho^0)^{d-2}}{\left(1+\tu^2\right)^{d-1}}
d\rho^0 \, d^{d-1} \tu
\end{split}
\end{equation}
corresponds to the chart \eq{xiparads}.
We define the field on the conformal boundary as
$\tphi(\tu)=\Rf{\varphi}(\rho)$. 
As in the EadS space, assuming further that on the 
horosphere \eq{hor:ds} the boundary field scales as 
$\tphi(p\tu)=p^{-\Delta}\tphi(\tu)$, we have 
\begin{equation}
\label{cft:assumds}
\begin{split}
\Rf{\varphi}(\rho) &= \Rf{\varphi}(\rho^0,\cdots,\rho^{d-1})\\
&=\tphi\left(
\frac{2 \tu^1}{1+\tilde{u}^2} \rho^0, 
\frac{2 \tu^2}{1+\tilde{u}^2} \rho^0, \cdots,
\frac{2 \tu^{d-1}}{1+\tilde{u}^2} \rho^0
\right) \\
&=\left(\frac{2\rho^0}{1+\tu^2}\right)^{-\Delta} 
\tphi(\tu).
\end{split}
\end{equation} 
Inserting this in \eq{iradds} and performing the integration over $\rho^0$
we obtain 
\begin{equation}
\label{ds2cone}
\varphi(U)=
\phi_0(d,\Delta)\int_{\sph{d-1}} {K}(t,u|\tu)
\tphi(\tu) \, d^{d-1}\tu \;,
\end{equation}
where the kernel is 
\begin{equation}
\label{smrfuncds}
{K}(t,u|\tu) = 
{\left(\frac{t}{-t^2+(u-\tu)^2}\right)^{d-1-\Delta}},
\end{equation} 
and $\phi_0(d,\Delta)$ is a constant,
\begin{equation}
\label{phi0:ds}
\phi_0(d,\Delta) = 
\frac{\cos\frac{(d-\Delta)\pi}{2}}{2^{\Delta-1}\pi^{d}e^{i\pi(d+\Delta)/2}
\cos^2\frac{d\pi}{2}}
\frac{\G{d/2}^2\G{1+\Delta}}{\G{\Delta+2-d}\G{(d-1)/2}^2}.
\end{equation}
Identifying $\varphi(U)$ with $\phi_1(U)$ or $\phi_2(U)$ in \eq{f:expand}
we obtain \eq{blk:rc} with the kernel \eq{smrfuncds} as before.
\subsubsection{Bulk field in the dS patch}
The free scalar field in the dS patch is obtained in the same way as in the
EadS patch, setting $|U|^2=1$ and plugging in \eq{Uvol}, \eq{UX:ds} and 
\eq{smrfuncds} in \eq{phi:fin} and performing the integration in $U=(t,u)$.
Similarly to the EadS patch, the Mellin transform of the bulk 
field with respect to the family parameter $\xi$ of the foliation 
\eq{patch:ds} is given by 
\begin{multline}
\label{Mf:ds}
\phi_s(\tau,x) = 
(1+(-1)^{s-1-\D})
\left(
\left(\frac{i}{2\pi}\right)^d 
(m)^{d-s} (2\tau)^s\G{s} 
\int_{\sph{d-1}} d^{d-1}\tu\tphi_1(\tu)
\I(-1,s,d-1-\Delta)\right. \\ \left.
+\left(\frac{i}{2\pi}\right)^d 
(-m)^{d-s} (2\tau)^s\G{s} 
\int_{\sph{d-1}} d^{d-1}\tu\tphi_2(\tu)
\I(-1,s,d-1-\Delta)
\right),
\end{multline}
with $\I$ as in \eq{master}. 
The expression is similar to the one for the EadS space, \eq{Mf:eads}. 
The first factor is new. This arises due to the difference of the domains
of integration in the two cases. Unlike the case of EadS space, 
the domain of $\xi$ in the dS space is 
the positive real line, but the timelike coordinate  $t$ in \eq{unit:ds}
is the whole real line. In order to match with the integral \eq{master} 
the $t$-integral  is to be split into
\begin{equation}
\int_{-\infty}^{\infty} dt = \int_{-\infty}^0 dt + \int_{0}^{\infty} dt.
\end{equation}
Changing $t\rt -t$ in the first integral and noting that only the factor
$t^{\alpha_1+\alpha_2-d}$ changes, since the other two factors contain only 
$t^2$, yields the first factor $(1+(-1)^{s-1-\D})$ of \eq{Mf:ds}.
We then use \eq{master} to write down the integrals for the dS space 
with $\epsilon=-1$. 
The expression is again of the form \eq{lin:com}, 
\begin{equation}
\I = C_1(\I_{123}^{(1)} + \I_{123}^{(2)}) + C_2\I_{234},
\end{equation}
with the three integrals given by the following  series. 
\begin{multline}
\label{ds:I123:1}
\I_{123}^{(1)}(-1,s,d-1-\Delta)=
-\sum_{n_1,n_2=0}^{\infty}
(-1)^{\D-s} 
2^{\frac{\D-2-s-2n_1}{2}}
\\
\frac{%
\sec\frac{\pi}{2}(1-2d+s+\D-2n_1-2n_2)\sin\pi(d-1-\D+2n_1+n_2)
\G{\frac{d-1}{2}}\G{\frac{s-\D+2n_1}{2}}
}{%
\scriptstyle
\G{1+s-\D-n_2}
\G{2-d+\frac{s+\D}{2}-n_1-n_2}
\G{\frac{d+1-2s}{2}+n_1+n_2}
\G{d-1-\D}\G{\D-s}\G{s}\G{1+n_1}\G{1+n_2}
}
\\ 
\tau^{d-1-2s+2n_1+2n_2}
(\tau^2-(x-\tu)^2)^{1-d+\frac{s+\D}{2}-n_1-n_2}
\pFq{\frac{s-\D}{2}+n_1,\frac{1-d}{2}+s-n_1-n_2}{
{\frac{4-2d+s+\D-2n_1-2n_2}{2}}}{\frac{\tau^2-(x-\tu)^2}{2\tau^2}},
\end{multline}
\begin{multline}
\label{ds:I123:2}
\I_{123}^{(2)}(-1,s,d-1-\Delta)=
(-1)^{\D-s}
\sum_{n_1,n_2=0}^{\infty}
2^{\D-d-2n_1-n_2}
\\
\frac{%
\cos\frac{\pi}{2}(d-2s+2n_1+2n_2)\sec\frac{\pi}{2}(1-2d+s+\D-2n_1-2n_2)
\G{\frac{d-1}{2}}
\G{\frac{d-1+s-\D}{2}}
}{%
\scriptstyle
\G{1+s-\D-n_2}
\G{2-d+\D-2n_1-n_2}
\G{d-\frac{s+\D}{2}+n_1+n_2}
\G{d-1-\D}\G{\D-s}\G{s}\G{1+n_1}\G{1+n_2}
}
\\
\tau^{1-d-s+\D}
\pFq{\frac{d-1-\D+s}{2},d-1-\D+2n_1+n_2}{d-\frac{s+\D}{2}+n_1+n_2}{
\frac{\tau^2-(x-\tu)^2}{2\tau^2}},
\end{multline}
\begin{multline}
\label{ds:I234}
\I_{234}(-1,s,d-1-\Delta)=
(-1)^{\D-s}
\sum_{n_1,n_2=0}^{\infty}
2^{\frac{d-3-3s+\D}{2}-n_1+n_2}
\\
\frac{%
\csc\frac{\pi}{2}(d-1+s-\D+2n_1)\sin\pi(1-2s+\D-2n_1+n_2)
\G{\frac{d-1}{2}}
\G{\frac{1-d+3s-\D+2n_1-2n_2}{2}}
}{%
\scriptstyle
\G{\frac{3-d-s+\D-2n_1}{2}}
\G{\frac{3-d}{2}+s+n_1-n_2}
\G{1+s-\D-n_2}
\G{d-1-\D}\G{\D-s}\G{s}\G{1+n_1}\G{1+n_2}
}%
\\
\tau^{2n_1}
(\tau^2-(x-\tu)^2)^{\frac{1-d-s+\D-2n_1}{2}}
\pFq{-n_1,\frac{1-d+3s-\D+2n_1-2n_2}{2}}{\frac{3-d-s+\D-2n_1}{2}}{
\frac{\tau^2-(x-\tu)^2}{2\tau^2}}.
\end{multline}
\section{The limit of massless scalar}
\label{sec:massless}
We have considered a massive scalar field in the Minkowski space. As
in the case of a celestial conformal field theory, a massless scalar is
not amenable to a similar description. If the mass $m$ of the scalar is
zero in \eq{X:Lap}, then, by \eq{eq:fR}, the second derivative of the Radon
transform $\Rf{\phi}$ of the field with respect to $\lambda$ vanishes. 
The inversion \eq{IRadon} is then singular, since it involves $d$ derivatives 
of $\Rf{\phi}$.

However, an expression for the bulk field in the limit of vanishing mass
can be obtained from the expressions derived above. While singular, 
it yields a solution to the boundary-value problem of the zero mode of the
Laplace operator in the Minkowski space, \emph{cum grano salis}. Let
us demonstrate this by taking the limit of vanishing mass in \eq{phi:fin}. It
suffices to write either term.

From the expression \eq{phi:fin}, we have
\begin{equation}
\label{mas0}
\begin{split}
\phi(X) &= \left(\frac{i}{2\pi}\right)^{d} \int\limits_{\dM} \dU
(m\sqrt{|U|^2})^d 
e^{-m\sqrt{|U|^2} U\cdot X} 
\int\limits_{\sph{d-1}} K(U| \tilde{u}) \tilde{\phi}_1(\tu) 
d^{d-1}\tu 
\\
&= \left(\frac{i\sqrt{\epsilon}}{2\pi}\right)^{d} 
\int\limits d^{d-1}u\ d^{d-1}\tu \ {dt}\left(\frac{m}{t}\right)^d
e^{\frac{\sqrt{-\epsilon} \xi}{2\tau}\left(\frac{m}{t}\right)
\big((u-x)^2+\epsilon\tau^2+\epsilon t^2\big)}
\left(\frac{t}{(u-\tu)^2-\epsilon t^2}\right)^{d-1-\D}\ \tilde{\phi}(\tu),
\end{split}
\end{equation}
where we used \eq{UX:eads}, \eq{UX:ds}, \eq{Uvol}, \eq{smrfunceads} and 
\eq{smrfuncds} in the second line and 
combined the expressions for the EadS and dS spaces by defining $\epsilon$
as in \eq{eps:def}.
In order to effect the singular limit of vanishing mass, we change variable 
\cite{Pasterski:2017kqt} in the integration from $t$ to 
\begin{equation}
y = \frac{m}{t}
\end{equation}
to write 
\begin{equation}
\phi(X)=m^{d-\D} \left(\frac{i\sqrt{-\epsilon}}{2\pi}\right)^{d} 
\int\limits d^{d-1}u\ d^{d-1}\tu \ y^{d-2}dy
e^{\frac{\sqrt{-\epsilon} \xi}{2\tau} y\big((u-x)^2+\epsilon\tau^2
+\tfrac{\epsilon m^2}{y^2} \big)}
\left(\frac{1/y}{(u-\tu)^2-\tfrac{\epsilon m^2}{y^2} }\right)^{d-1-\D}\! \tilde{\phi}(\tu).
\end{equation}
The limit $m\rt 0$ is ill-defined due to the leading factor of $m^{d-\D}$.
Taking the limit only in the integrand we then have
\begin{equation}
\phi(X) \stackrel{m\rt 0}{\sim} m^{d-\D} 
\left(\frac{i\sqrt{-\epsilon}}{2\pi}\right)^{d} 
\int\limits d^{d-1}u\ d^{d-1}\tu \ y^{d-2}dy
e^{\frac{\sqrt{-\epsilon} \xi}{2\tau} y\big((u-x)^2+\epsilon\tau^2\big)}
\left(\frac{1}{y(u-\tu)^2}\right)^{d-1-\D}\ \tilde{\phi}(\tu).
\end{equation}
Changing variable further from $y$ to 
\begin{equation}
y' = 
\frac{\sqrt{-\epsilon} \xi}{2\tau} y\big((u-x)^2+\epsilon\tau^2\big),
\end{equation}
we derive
\begin{equation}
\phi(X) \stackrel{m\rt 0}{\sim} m^{d-\D} 
\left(\frac{i\sqrt{-\epsilon}}{2\pi}\right)^{d} 
\left(\frac{2\tau}{\sqrt{-\epsilon}\xi}\right)^\D
\int (y')^{\D-1}e^{y'}dy'
\int \frac{d^{d-1}u}{\big((u-x)^2+\epsilon\tau^2\big)^\D}
\int \frac{d^{d-1}\tu\ \tilde{\phi}(\tu)}{(\tu-u)^{2(d-1-\D)}}.
\end{equation}
The integration over $y'$ is a constant, possibly divergent. 
Since the field $\tilde{\phi}(\tu)$ is conformal by assumption, 
it does not change under
a shift of $\tu$ by $u$ in the integral. 
In the second integral, $u$ can be shifted by $x$ as well. Ignoring all the
constants, including $m$, we then obtain 
\begin{equation}
\label{phiX:m0}
\phi(X) \sim \frac{\tau^{d-1-\D}}{\xi^\D}
\int \frac{d^{d-1}\tu\ \tilde{\phi}(\tu)}{\tu^{2(d-1-\D)}}.
\end{equation}
Let us note that bulk field does not depend on $x$. The integration over the
field $\tilde{\phi}$ yields a constant. 
The field $\phi(X)$ satisfies the Laplace equation in the Minkowski space
\eq{X:Lap} with $m=0$, which can be verified by writing the Laplacian in
the appropriate chart as
\begin{equation}
\square =\frac{\tau^2}{\xi^2}\nabla_x^2+\frac{1}{\xi^2}
 \qty({\tau^2}\partial_\tau^2 - {(d-2)\tau} \partial_\tau) - \partial_\xi^2 - \frac{d}{\xi} \partial_\xi.
\end{equation}
Hence, the limiting expression \eq{phiX:m0} solves a boundary value problem
of the Laplace equation in the Minkowski space with the boundary value
given by the integral of the field $\tilde{\phi}$.
\section{Perspective from representation theory}
\label{sec:rep}
Let us briefly discuss the construction presented here in the light of 
representation theory of orthogonal groups. Let $G=SO_0(1,d)$ denote the 
proper Lorentz group which acts on $X\in\Md$ as 
$(d+1)\times (d+1)$ matrices  preserving the norm \eq{Mnorm}.
Let us define
\begin{gather}
    M = \left\{\begin{pmatrix}
    1&0&0\\
    0&m&0\\
    0&0&1
    \end{pmatrix}; \quad m\in SO(d-1)\right\},
    \\
    A=\left\{\begin{pmatrix}
        \cosh\alpha &0& \sinh\alpha\\
        0&I_{d-1}&0\\
        \sinh\alpha &0& \cosh\alpha
    \end{pmatrix},\quad\alpha\in\R\right\},
    \\
    N=\left\{\begin{pmatrix}
        1+\frac{x^2}{2\tau^2} & x^1 &\cdots &x^d & -\frac{x^2}{2\tau^2}\\
        x^1 & & & &-x^1\\
        x^2 & & & &-x^2\\
        \vdots &&I_{d-1}&& \vdots\\
        x^d & & & &-x^d\\
        \frac{x^2}{2\tau^2} & x^1 &\cdots &x^d & 1-\frac{x^2}{2\tau^2}
    \end{pmatrix}; {x^i\in\R\atop i=1,2,\cdots,d} \right\},
\end{gather}
where we used the charts \eq{patch:eads} and \eq{patch:ds} to write $N$. 
Then  $P=MAN$, known as the  Langlands decomposition of $G$,
is the isotropy group of the point
$X_\star=(X^0,X^i,X^d)=(1,0,1)$ on the light cone. 
Since $G$ acts transitively on $\Md$, the orbit of $X_\star$ under 
$P$ in $G$ is the sphere $\sph{d-1}$ obtained as the homogeneous
space $G/P$. The sphere is obtained by first foliating $\Md$ with EadS or dS 
spaces using the hyperplanes \eq{u1} and then taking their boundaries on the
light cone using \eq{hor:eads} and \eq{hor:ds}. 
In terms of coordinate charts, the light cone $|X|^2=0$ is obtained as
$\tau\rt0$, yielding 
\begin{equation}
X^0 = \pm\frac{\xi}{2\tau}  (1+x^2), \quad X^i = \frac{x^i \xi}{\tau},  
	\quad X^d = \mp\frac{\xi}{2\tau}  (1-x^2),
\end{equation}
for \eq{patch:eads} and \eq{patch:ds}, respectively. 
Setting $X^0=\rho^0$ leads to the coordinates \eq{xipara} and \eq{xiparads}, 
respectively. The unit sphere inside the light cone is given by 
$\sum\limits_{i=1}^d (X^i)^2=(X^0)^2=1$, with parametrisations following from
\eq{xipara} and \eq{xiparads}.

We have dealt with scalar fields here, namely, $\phi(X)$ on $\Md$, and 
$\tilde{\phi}(\tu)$ on $\sph{d-1}$. Generally, 
representations of $G$ in the Minkowski spacetime, 
{\sl e.g.}, the principal series $(\sigma,\nu)$, is obtained by 
parabolic induction \cite{Sun} using the parabolic subgroup $P$ by
picking an irreducible representation $\sigma$ of the compact group $M$ 
and a character of the maximal torus $A$ as
\begin{equation}
	\label{rep1}
	\phi=T_g(\tilde{\phi})= a^{\nu+\frac{d-1}{2}} \sigma(m)
	\tilde{\phi}(g man),\quad g\in G,m\in M,
    a\in A, n\in N.
\end{equation}
This is the relation between the conformal field $\tilde{\phi}$ on the sphere
$\sph{d-1}$  and the field on $\Md$, valued in the principal series 
representation of the proper Lorentz group. 
$\nu$ is related to the conformal dimension $\D$ of $\tilde{\phi}(\tu)$.
The integrations in \eq{phi:fin} are averaging over the parabolic group, 
as is customary in integral geometry. 

\section{Summary and discussions}
\label{sec:summary}
In this article, we obtain a correspondence between a  
free scalar field of mass $m$ in the bulk 
$(d+1)$-dimensional Minkowski space and a scalar field 
with scaling dimension $\D$  on a 
$(d-1)$-dimensional sphere in it. The correspondence, while similar 
in spirit to holography, is devoid of gravitational content.
The construction proceeds in two steps. The first step is to  Radon
transform the scalar field in the bulk to hyperplanes determined by 
unit vectors in the Minkowski space. The family of such hyperplanes covers
different regions of the bulk, as depicted in Figure~\ref{fig1}. A
unit vector determines an EadS or a dS slice. The importance of taking the
Radon transform lies in the fact that it allows separating the field into
two parts, one depending on the family parameter and the other on the
EadS or dS slice. The latter is then interpreted as a scalar field on the
EadS or dS space and is taken to have been obtained from a scalar field
of scaling dimension $\Delta$ on a sphere on its boundary through the HKLL 
bulk reconstruction. The sphere is different from the celestial sphere in that
it does not reside in the null hypersurface of the bulk. Rather, it forms
the boundary of the EadS or dS slice. 
The construction  utilises the holographic bulk reconstruction in an
essential manner and is described in terms of invertible integral transforms 
only. We write down the Mellin modes of the scalar field in the bulk. 
Noting that the kernels appearing 
in the integrations involve only quadratic forms of the coordinates, 
as for Feynman loop diagrams, we use the Lee-Pomeransky method to
evaluate the integrals. The Mellin modes are expressed in terms of GKZ
hypergeometric series, exhibiting a rich structure.

An apparent shortcoming of the present
construction is that the constants $\phi_0$ in \eq{phi0:eads} and 
\eq{phi0:ds} are vanishing and singular, respectively, for odd $d$. 
However, since the troublesome factors depend on the dimension of the 
spacetime alone, this can be 
taken care of by normalising the fields by dividing by such factors without
affecting physical considerations. 

Thought of as a  boundary-value problem 
it is natural to look askance at any correspondence that attempts to relate
fields satisfying linear differential equations as in here 
on spaces whose dimensions differ by two.
\footnote{KR thanks Gautam Mandal for a useful discussion on this issue} 
The functions 
${\phi}_1$ and ${\phi}_2$ in \eq{f:expand} are arbitrary functions
of $U$ arising from the oscillator equation \eq{eq:fR}. Constants in 
the linear combination \eq{f:expand} can be fixed by demanding certain
asymptotic behaviour of $\Rf{\phi}$ with respect to $\lambda$. This does not
fix the ${\phi}$'s. In the second step of the correspondence, in \eq{blk:rc},
we \emph{identify} these with the field reconstructed from the sphere, 
as in \eq{eads2cone} or \eq{ds2cone}. These in turn satisfy Laplace equations
in the respective EadS or dS spaces, namely,
\begin{equation}
\label{shado}
\Box_U\varphi(U) = \epsilon \D(\D-d+1)\varphi(U),
\end{equation}
with $\epsilon=1$ for the EadS space and $\epsilon=-1$ for the dS space, 
with $\Box_U$ denoting the Laplacian in the respective spaces
\cite{Bhowmick:2017gj,Bhowmick:2019nso}.
Further constants are to be fixed in
choosing the solution of this Laplace equation, thereby making the 
correspondence consistent. Furthermore, since the Laplace equation 
\eq{shado} remains
unaltered under $\D \rt (d-\D-1)$, the scaling dimension of the 
field $\varphi$ may be taken to be $(d-\D-1)$ instead of $\D$, 
corresponding to 
the so-called shadow field. This does not alter the identifications made 
above. 

The computations do not rigorously extend to massless fields in the Minkowski 
space. However, ignoring the singular parts, it solves a boundary value
problem of the Laplace equation, with the boundary value specified 
by the field on $\sph{d-1}$ integrated over the sphere. 

We have also discussed the relationship of the integral geometry construction
presented here with group representation theory, showing how the conformal
fields on the sphere $\sph{d-1}$ relates to the principal series representation
of the proper Lorentz group of $\Md$ using parabolic induction. 

The correspondence can be extended to transforming correlation functions of
a conformal field theory on the sphere to the Minkowski bulk using similar 
methods, resulting in GKZ hypergeometric functions again. 
This will be reported in future. We hope that the present article will be 
useful in understanding the cherished flat space holography.

\section*{Acknowledgement}
KR thanks Shankhadeep Chakraborty, 
Diptarka Das, Apratim Kaviraj and  Gautam Mandal for useful discussions.

\appendix
\renewcommand{\theequation}{A.\arabic{equation}}
\renewcommand{\theHequation}{A.\arabic{equation}}
\renewcommand{\thesubsection}{A:\Roman{subsection}}

\setcounter{equation}{0}
\section{Evaluation of $\I$ using Lee-Pomeransky method}
In this appendix we detail the evaluation of the integral 
\begin{equation}
\label{master}
\I(\epsilon,\alpha_1,\alpha_2) = \int_0^{\infty} dt\ 
t^{\alpha_1+\alpha_2-d} \int\frac{d^{d-1}u}{%
\big(\epsilon t^2+\epsilon
\tau^2+(x-u)^2\big)^{\alpha_1}\big(\epsilon t^2+(u-\tu)^2\big)^{\alpha_2}}
\end{equation}
appearing in \eq{Mf:eads} and \eq{Mf:ds}. The $(d-1)$-dimensional integral 
over $u$ is performed using the Lee-Pomeransky 
method originally formulated for evaluation of Feynman integrals. 
The parameter $\epsilon$ is to be chosen according to
\begin{equation}
\label{eps:def}
\epsilon=\begin{cases}
-1, \quad \text{for dS},\ \ds\\
\phantom{-} 1, \quad \text{for EadS},\ \eads.
\end{cases}
\end{equation}
\subsection{Lee-Pomeransky method}
We briefly describe the combinatorial aspects of the method of 
Lee-Pomeransky \cite{klau,wein,lee}.
The goal is to evaluate the integral over $L$ vectors 
$\u_i\in\R^D$, of the form
\begin{equation}
\label{feyn}
I(\alpha_1,\cdots,\alpha_n) = \int 
\frac{d^D\u_1\ldots d^D\u_L}{G_1^{\alpha_1}\ldots G_n^{\alpha_n}},
\end{equation}
for an $n$-tuple $\boldsymbol\alpha = (\alpha_1,\alpha_2,\ldots,\alpha_n)$
and $n$ quadratic forms $G_k$ in $\u_i$, with $k=1,2,\ldots,n$
and $i=1,2,\ldots,L$.
Introducing a set of real-valued parameters 
$\eta_1,\ldots,\eta_n$, and an $L$-tuple of vectors 
$v=(\u_1,\ldots,\u_L)$, one writes the quadratic form in $\u$ as
\begin{equation}
\eta_1G_1+\cdots+\eta_nG_n = v^t A v +2 B^t v + C,
\end{equation}
where $A$ is an $L\times L$ matrix, $B$ is an $L$-component column matrix, and
$C$ is a scalar, all of which are independent of $\u$, while depend on the 
$\eta$'s. Here a superscript $t$
denotes the transpose of a matrix and the products on the RHS are taken to
be dot products of vectors $\u$. The Symanzick polynomials are then defined as
\begin{equation}
\label{syman}
\begin{gathered}
U = \det A\\
F = (B^t A^{-1} B - C)U,
\end{gathered}
\end{equation}
which in turn are used to define the Lee-Pomeransky polynomial 
\begin{equation}
\label{Gpol}
G = F+U.
\end{equation}
The integral \eq{feyn} is then rewritten as 
\begin{equation}
\label{leepom}
I(\alpha_1,\cdots,\alpha_n) = \frac{(-1)^{|\boldsymbol\alpha|}\G{D/2}}{%
\G{{(1+L)D}/{2}-|\boldsymbol\alpha|}\G{\alpha_1}\ldots\G{\alpha_n}}
\int\limits_0^{\infty}\!\! 
\cdots\!\!\!
\int\limits_0^{\infty} 
d\eta_1\ldots d\eta_n 
\eta_1^{\alpha_1-1} \ldots \eta_n^{\alpha_n-1} G^{-D/2},
\end{equation}
for $\alpha_k>0$, for all $k=1,\ldots,n$ and we defined 
$|\boldsymbol\alpha|=\alpha_1+\cdots+\alpha_n$.

The integral is evaluated as a generalised GKZ hypergeometric function using
combinatorial data extracted from the Lee-Pomeransky polynomial. The 
polynomial $G$ is a sum of $N$ monomials in powers of $\eta$'s, where 
$N$ depends on the specific integral to be evaluated. 
Denoting the coefficients of the monomials as $z$, the polynomial is
written as 
\begin{equation}
\label{mono}
G = \sum\limits_{j=1}^N z_j \eta_1^{a_{1j}}\ldots \eta_n^{a_{nj}}, 
\end{equation}
where the positive  integer $a_{kj}$ denotes the index of $\eta_k$ in the 
$j$-th monomials in $G$, written in a fixed order, with $k=1,\ldots,n$ 
and $j=1,\ldots,N$. They are arranged into an $n\times N$ 
rectangular matrix with columns given by $n$-dimensional vectors. These
vectors possess an affine structure. Thus, it is convenient to write them
in a projective setting by adding a row of unity to the rectangular matrix
defining the $(n+1)\times N$ matrix of positive integers
\begin{equation}
\label{toricA}
A = \begin{pmatrix}
\ba_1 & \ba_2  \cdots \ba_N
\end{pmatrix} = \bordermatrix{%
& z_1 & z_2 & \cdots & z_N \cr
& 1   &  1  & \cdots & 1 \cr
& a_{11} & a_{21} & \cdots & a_{N1}\cr
& a_{12} & a_{22} & \cdots & a_{N2}\cr
& \vdots & \vdots & \ddots & \vdots \cr
& a_{1n} & a_{2n} & \cdots & a_{Nn}
}%
\in \Z_{\geqslant 0}^{(n+1)\times N},
\end{equation}
which we refer to as the toric matrix. Each column corresponds to a monomial
of $G$. The coefficient of a monomial in \eq{mono} 
is indicated on top of each column. Let us point out the 
unconventional indexing of rows and columns in the toric matrix,
with $a_{ij}$ denoting the entry in the $j$-th row and $i$-th column. 
We shall have occasions to choose from
among the columns of the toric matrix. For an index set 
$\sigma\subset \{1,2,\ldots,N\}$, we use the notation
$A_{\sigma}$ to indicate the submatrix obtained from the columns of $A$ 
indexed by the elements of $\sigma$. Similarly, $z_{\sigma}$ is taken
to define the set of $z$'s indexed by $\sigma$. 

For an index set $\sigma \subset \{1,2,\ldots,N\}$, the positive span of 
the column vectors, indexed by $\sigma$ is called a cone, that is,
\begin{equation}
\cone{\sigma} = \sum_{I\in\sigma} \R_{\geqslant 0}\ba_I.
\end{equation}
The subset $\sigma$, the subset of vectors $\{\ba_I|\ I\in\sigma\}$ and
$\cone{\sigma}$ are used to refer to each other interchangeably. 
The complement of $\sigma$ in $\{1,2,\ldots,N\}$ is denoted $\bsigma$.
A subset $\mathcal{T}$ of the power set of $\{1,2,\ldots,N\}$ is called a 
triangulation if 
$\{\cone{\sigma}|\sigma\in \mathcal{T}\}$ is the set of cones in 
a simplicial fan with 
support equal to $\cone{A}$. For an $(n+1)$-dimensional real vector $w$, 
a triangulation $\mathcal{T}(w)$ is defined if there exists an 
$(n+1)$-dimensional real vector $\mathbf{m}$ such that  
\begin{equation}
\mathbf{m}\cdot\ba_I  \begin{cases}
=w_I,\quad \text{if}\ I\in\sigma,\\
< w_I, \quad\text{if}\ I\in\bsigma.
\end{cases}
\end{equation}
A triangulation is regular if $\mathcal{T}=\mathcal{T}(w)$ for some $w$.

Given a regular triangulation $\mathcal{T}$ 
corresponding to the columns of $A$, let 
$\sA$ denote the $(n+1)\times (n+1)$ matrix made of $\{\ba_I\}$, 
$I\in\sigma$ and 
$\bA$ denote the $(n+1)\times (N-n-1)$ matrix made of the vectors $\{\ba_I\}$, 
$I\in\bsigma$. By rearranging the columns, as required, the toric  matrix
$A$ can be expressed as $A=(\sA|\bA)$ with the columns reshuffled.
If $\sigma$ belongs to a regular triangulation, then $|\det \sA|=1$.
The integral is given by the $\Gamma$-series \cite{heo,PR3}, 
\begin{equation}
\label{G-ser}
\Psi_{\sigma} = 
\prod\limits_{I\in\sigma}
z_I^{-\left(\sA^{-1}\m\right)_I}
\sum_{\mathbf{n}\in\Z_{\geq 0}^{N-n-1}} \frac{%
\prod\limits_{I\in\sigma}
z_I^{-\left(\sA^{-1}\bA\mathbf{n}\right)_I}
\prod\limits_{J\in\bsigma}
z_J^{{n}_J}}%
{%
\prod\limits_{I\in\sigma}
\G{1-\left(\sA^{-1}\m\right)_I-\left(\sA^{-1}\bA\mathbf{n}\right)_I}
\prod\limits_{J=1}^{N-n-1}\G{1+n_J}
},
\end{equation}
where $(\mathbf{v})_I={v}_I$ is taken to denote the $I$-th components 
of a vector $\mathbf{v}$ and the $n$-tuple $\boldsymbol\alpha$ is extended to 
$\m = (D/2,\alpha_1,\alpha_2,\ldots,\alpha_n)$.
The integral \eq{leepom} is a linear combination of series corresponding to
all the cones of a regular triangulation of \eq{toricA}.
\subsection{The integral \eq{master}}
The integral \eq{master} is evaluated in two steps. The $(d-1)$-dimensional
integral over $u$ is performed using the Lee-Pomeransky method, followed by
the integration over $t$.

Let us discuss the first step. The integral over $u$ is of the form \eq{feyn}
with $L=1$, $D=d-1$, $n=2$ and 
\begin{gather}
G_1 =  \epsilon t^2+\epsilon \tau^2+(x-u)^2\\
G_2 =  \epsilon t^2+(u-\tu)^2.
\end{gather} 
From these, the toric matrix \eq{toricA} is evaluated to be 
\begin{equation}
\label{torMat}
 A = \bordermatrix{ 
  & -\epsilon(t^2+\tau^2) & -\epsilon(2t^2+\tau^2)-(x-\tilde{u})^2 & 
  1 & -\epsilon t^2 & 1 \cr 
  &1 & 1 & 1 & 1 & 1 \cr
 &2 & 1 & 1 & 0 & 0 \cr
 &0 & 1 & 0 & 2 & 1       },
     \end{equation}
with the coefficients $(z_1,\ldots,z_5)$ indicated above the columns.
The cones of a regular triangulation are
$\{1,2,3\}$, $\{2,3,4\}$, $\{3,4,5\}$.
Correspondingly, the three series are 
{\tiny
\begin{align}
\psi_{\{1,2,3\}} &= \sum_{n_1,n_2=1}^\infty
\frac{i^{d-1} 
\epsilon^{\frac{1}{2}\left(d-1-2 \alpha_1+4 n_1+2 n_2\right)} 
t^{2n_1} 
\left(t^2+\tau ^2\right)^{\tfrac{d-1}{2}-\alpha_1+n_1+n_2} 
\left(\left(x-\tu\right)^2+\epsilon\left(
2 t^2+\tau ^2\right)\right)^{-\alpha_2 -2 n_1-n_2}
\G{\frac{d-1}{2}} }{%
\G{n_1\!+\!1}\G{n_2\!+\!1} 
\G{1\!-\!\alpha_2\!-\!2 n_1\!-\!n_2} 
\G{\tfrac{d\!+\!1}{2}\!-\!\alpha_1\!+\!n_1\!+\!n_2}
\G{2\!-\!d\!+\!\alpha_1\!+\!\alpha_2\!-\!n_2}
\G{d\!-\!1\!-\!\alpha_1\!-\!\alpha_2}\G{\alpha_1}\G{\alpha_2}
} \\ %
\psi_{\{2,3,4\}} &= \sum_{n_1,n_2=1}^\infty
\frac{i^{d-1} 
\epsilon^{\frac{1}{2}(1-d+2\alpha_1+4n_1-2n_2)}
t^{1-d+2\alpha_1+2 n_1-2 n_2}
\left(t^2+\tau ^2\right)^{n_1} 
   \left(\left(x-\tu\right)^2+\epsilon  \left(2 t^2+\tau ^2\right)\right)^{d-1-2 \alpha_1  -\alpha_2-2 n_1+n_2}\G{\frac{d-1}{2}}}{%
\G{n_1\!+\!1}\G{n_2\!+\!1}
\G{\frac{3-d}{2}+\alpha_1+n_1-n_2} 
\G{2\!-\!d\!+\!\alpha_1\!+\!\alpha_2\!-\!n_2} 
\G{d-2 \alpha_1 -\alpha_2-2 n_1+n_2}
\G{d\!-\!1\!-\!\alpha_1\!-\!\alpha_2}\G{\alpha_1}\G{\alpha_2}
}\\%
\psi_{\{3,4,5\}} &= \sum_{n_1,n_2=1}^\infty
\frac{i^{d-1} 
\epsilon^{\frac{1}{2}(d-1-2\alpha_1-2\alpha_2-2 n_2)} 
t^{d-1-2\alpha_1-2\alpha_2 -2n_1-2n_2}
\left(t^2+\tau ^2\right)^{n_1}  \left(\left(x-\tu\right)^2+\epsilon  
\left(2 t^2+\tau ^2\right)\right)^{n_2}\G{\frac{d-1}{2}}}{%
\G{n_1\!+\!1}\G{n_2\!+\!1}
\G{1\!-\!\alpha_1\!-\!2 n_1\!-\!n_2}
\G{\tfrac{d\!+\!1}{2}\!-\!\alpha_1\!-\!\alpha_2\!-\!n_1\!-\!n_2}
\G{2\!-\!d+2 \alpha_1\!+\!\alpha_2\!+\!2n_1\!+\!n_2}
\G{d\!-\!1\!-\!\alpha_1\!-\!\alpha_2}\G{\alpha_1}\G{\alpha_2}
}%
\end{align}
}
Inserting these in \eq{master} and performing the integration over $t$ after
multiplying by $t^{\alpha_1+\alpha_2-d}$, we obtain the integral 
\eq{master}  as a linear combination of contribution from the three cones,
$\sigma\in\mathcal{T}=\{\{1,2,3\},\{2,3,4\},\{3,4,5\}\}$, with
\begin{equation}
\I_{\sigma}(\epsilon,\alpha_1,\alpha_2) = 
\sum_{n_1,n_2 = 0}^\infty 
\I_{\sigma}^{(1)}(n_1,n_2) + 
\sum_{n_1,n_2 = 0}^\infty 
\I_{\sigma}^{(2)}(n_1,n_2),
\end{equation}
where $\I_{\sigma}^{(1)}(n_1,n_2)$ and $\I_{\sigma}^{(2)}(n_1,n_2)$ 
are series given by 
\begin{multline}
\label{gen:I123:1}
\I_{\{1,2,3\}}^{(1)}(n_1,n_2) = 
i^{d+1} 
\epsilon^{d-1 + n_{1} + n_{2} - \tfrac{3\alpha_{1} + \alpha_{2}}{2}} 
2^{\tfrac{\!d -3  - 2n_{1} - \alpha_{1} - \alpha_{2}}{2}} 
\\ \times 
\tfrac{\sec(\frac{\pi}{2}(d + 2n_{1} + 2n_{2} - \alpha_{1} + \alpha_{2})) 
\sin((2n_{1} + n_{2} + \alpha_{2})\pi)
\G{\tfrac{1}{2}(1 - d + 2n_{1} + \alpha_{1} + \alpha_{2})}
}{
\G{\tfrac{1}{2} + \tfrac{d}{2} + n_{1} + n_{2} - \alpha_{1}} 
\G{\tfrac{1}{2}(3 - d - 2n_{1} - 2n_{2} + \alpha_{1} - \alpha_{2})} 
\G{2 - d - n_{2} + \alpha_{1} + \alpha_{2}} 
\G{1+n_{1}}\G{1+n_{2}}
}
\tfrac{\G{\frac{d-1}{2}}}{\G{\alpha_1}\G{\alpha_2}\G{d-1-\alpha_1-\alpha_2}}
\\
 \times 
\tau^{d -1  + 2(n_{1} + n_{2} - \alpha_{1})}
((x - \tu)^{2}+\epsilon\tau^2)^{\tfrac{1 - d - 2(n_{1} + n_{2}) + \alpha_{1} - \alpha_{2}}{2}}
   \pFq{
       \tfrac{1}{2} - \tfrac{d}{2} - n_{1} - n_{2} + \alpha_{1}, 
       \tfrac{1}{2}\!\left(1 - d + 2n_{1} + \alpha_{1} + \alpha_{2}\right)}{
       \tfrac{1}{2}\!\left(3 - d - 2n_{1} - 2n_{2} + \alpha_{1} - \alpha_{2}\right)}{ 
       \tfrac{\epsilon\tau^{2} + (x - \tu)^{2}}{2\epsilon\tau^{2}}
   }
\end{multline},
and
\begin{multline}
\label{gen:I123:2}
\I_{\{1,2,3\}}^{(2)}(n_1,n_2) = 
-i^{d+1}
\epsilon^{ \frac{d-1}{2}- (\alpha_{1} +\alpha_{2})}  
2^{-(1 + 2n_{1} + n_{2} + \alpha_{2})}  
\\
\times 
\tfrac{
\cos(\tfrac{\pi}{2}(d + 2n_{1} + 2n_{2} - 2\alpha_{1}))  
\sec(\tfrac{\pi}{2}(d + 2n_{1} + 2n_{2} - \alpha_{1} + \alpha_{2}))
\G{\frac{\alpha_{1} + \alpha_{2}}{2}}
}{%
\G{1 - 2n_{1} - n_{2} - \alpha_{2}} 
\G{\tfrac{1}{2}(1 + d + 2n_{1} + 2n_{2} - \alpha_{1} + \alpha_{2})} 
\G{2 - d - n_{2} + \alpha_{1} + \alpha_{2}} 
\G{1+n_{1}} \G{1+n_{2}}}
\tfrac{\G{\frac{d-1}{2}}}{\G{\alpha_1}\G{\alpha_2}\G{d-1-\alpha_1-\alpha_2}}
\\
\times 
\tau^{-(\alpha_{1} + \alpha_{2})}
\pFq{2n_{1} + n_{2} + \alpha_{2},\tfrac{1}{2}(\alpha_{1} + \alpha_{2})}{
\tfrac{1}{2}(1 + d + 2n_{1} + 2n_{2} - \alpha_{1} + \alpha_{2})}{
\tfrac{\epsilon\tau^{2} + (x-\tu)^{2}}{2\epsilon\tau^{2}}
}
\end{multline}
for $\sigma=\{1,2,3\}$. The integrals arising from $\sigma=\{3,4,5\}$ and one
part of the integrals arising from $\sigma=\{2,3,4\}$ vanish due to the 
presence of singular 
factors of the form $\G{-n_1}$ or $\G{-n_2}$ in the denominators. 
For $\sigma = \{2,3,4\}$ the integral takes the form
\begin{multline}
\label{gen:I234}
\I_{\{2,3,4\}}(n_1,n_2) = 
i^{d+1}
\epsilon^{\tfrac{d-1-\alpha_1-\alpha_2}{2}+n_1}
2^{d-2-\tfrac{3\alpha_1+\alpha_2}{2}-n_1+n_2}
\\
 \times 
\tfrac{
\sin\pi(d-2\alpha_1-\alpha_2-2n_1+n_2)
\sec\pi(\tfrac{1+\alpha_1+\alpha_2+2n_1}{2})
\Gamma\!\qty({1-d+\tfrac{3\alpha_1+\alpha_2}{2}+n_1-n_2}) 
}{%
\Gamma\!\qty({1-\tfrac{\alpha_1+\alpha_2}{2}-n_1})\Gamma\!\qty({\tfrac{3-d}{2}+\alpha_1+n_1-n_2})
\Gamma\!\qty({2-d+\alpha_1+\alpha_2-n_2})
\Gamma\!\qty({n_1+1})\Gamma\!\qty({n_2+1})
}
\tfrac{\G{\frac{d-1}{2}}}{\G{\alpha_1}\G{\alpha_2}\G{d-1-\alpha_1-\alpha_2}}
\\
\times 
\tau^{2n_1}
\qty[(x-\tu)^2+\epsilon\tau^2]^{-\tfrac{\alpha_1+\alpha_2+2n_1}{2}}
\pFq{-n_1,\,1-d+\tfrac{3\alpha_1+\alpha_2}{2}+n_1-n_2}{
1-\tfrac{\alpha_1+\alpha_2}{2} - n_1}{\tfrac{(x-\tu)^2+\epsilon\tau^2}{2\epsilon\tau^2}}.
\end{multline}

\end{document}